# Geomagnetic effects on cosmic ray propagation under different conditions for Buenos Aires and Marambio, Argentina

**Jimmy J. Masías-Meza[1] and Sergio Dasso[2,3]**

[1] Departamento de Física (FCEN-UBA-IFIBA), Buenos Aires, Argentina

[2] Departamento de Física (FCEN-UBA), Buenos Aires, Argentina

[3] Instituto de Astronomía y Física del Espacio (UBA-CONICET), Buenos Aires, Argentina

email: (masiasmj@df.uba.ar).

*Abstract*. The geomagnetic field ($B_{geo}$) sets a lower cutoff rigidity ($R_c$) to the entry of cosmic particles to Earth which depends on the geomagnetic activity. From numerical simulations of the trajectory of a proton using different models for $B_{geo}$ (performed with the MAGCOS code), we use backtracking to analyze particles arriving at the location of two nodes of the net LAGO (Large Aperture Gamma ray burst Observatory) that will be built in the near future: Buenos Aires and Marambio (Antarctica), Argentina. We determine the asymptotic trajectories and the values of $R_c$ for different incidence directions, for each node. Simulations were done using several models for $B_{geo}$ that emulate different geomagnetic conditions. The presented results will help to make analysis of future observations of the flux of cosmic rays done at these two LAGO nodes.

## 1. Introduction

In the present work we report effects of the geomagnetic field on the arrival of low energy cosmic rays (CRs, primary particles with energies lower than ~100GeV) to two ground locations where new nodes of the Large Aperture Gamma ray burst Observatory (LAGO) will be constructed in the near future, one in Buenos Aires and the another in the Marambio base of Antarctica, both in Argentina. The LAGO project [1] aims at observing Gamma Ray Bursts (GRBs) by the single particle technique using water Cherenkov detectors. These detectors can be also used to study the Galactic Cosmic Ray flux at Earth.

In particular, we make numerical simulations of the trajectory of a proton and analyze the main properties of the arrival at these locations, such as asymptotic trajectories and values for the rigidity cutoff ($R_c$) for different incidence directions using the MAGCOS code (http://cosray.unibe.ch/~laurent/magnetocosmics).

Similar studies to the one presented here have been made for the site of the Pierre Auger Observatory, at Malargüe, Argentina [2].

The International Geomagnetic Reference Field (IGRF, see [3]) is a semi empirical description of the Earth's magnetic field (until ~ 5 $R_E$ from the center of the Earth), updated every 5 years since 1955, and supported from data provided by satellites, observatories and surveys around the world. This model is mainly of dipolar topology and includes the secular variation of the main dipole moment, the angular displacement of the geomagnetic axis respect to the Earth rotation axis, and the spatial displacement of the dipole location from the Earth's center. In figure 1a, we show the magnetic topology at the meridional plane that contains the direction to the Sun corresponding to a centered dipolar field (green) and the IGRF model (red). The latest model is valid until the year 2015 [3].

The geomagnetic field can be given by $B_{geo} = -\nabla V$, with:

$$V(r, \theta, \phi, t) = $$
$$= R_E \sum_{n=1}^{N} \left(\frac{R_E}{r}\right)^{n+1} \sum_{m=0}^{n} [g_n^m(t) \cos(m\phi) + $$
$$+ h_n^m(t) \sin(m\phi)] P_n^m(\cos\theta)$$

where $r$ is the radial distance from the center of the Earth, $\theta$ is the geocentric co-latitud, $\phi$ is the east longitude measured from the Greenwich meridian, $P^m_n$ are the Schmidt semi-normalized associated Legendre functions of degree $n$ and order $m$, $g(t)$ and $h(t)$ are fitted time-dependent coefficients [3].

On the other hand, some effects of the solar wind on the main magnetospheric current systems (e.g. the azimutal ring current, magnetotail currents, magnetopause and other field-aligned currents or Birkeland currents) can be modeled from observations of the solar wind dynamic pressure (Pdyn) using an advanced model [4], which includes the magnetic configuration of the magnetosphere for calm and for active conditions (e.g., geomagnetic storm). In the present work, we use the Tsyganenko 2001 (TSY01) model version [4] for describing the effects of the solar wind on the magnetic configuration of the outer magnetosphere.

In figure 1b, we show magnetic field measurements by the spacecraft Explorer XII, superposed with a line that represents the spatial distribution of a magnetic dipole. We see that until ~5$R_E$, th geomagnetic field is approximately dipolar.

In figure 1c we show the dipolar field (green) and the TSY01 model (red), where the effect of the solar wind dynamic pressure on the position of the magnetopause in the day side is clearly seen.

It is worth to note that this model includes the IGRF model and is only valid inside the magnetosphere; that is, TSY01 is valid from the Earth ground level until the magnetopause (modeled by a paraboloid with its main largest axis in the Earth-Sun direction, see [4]).

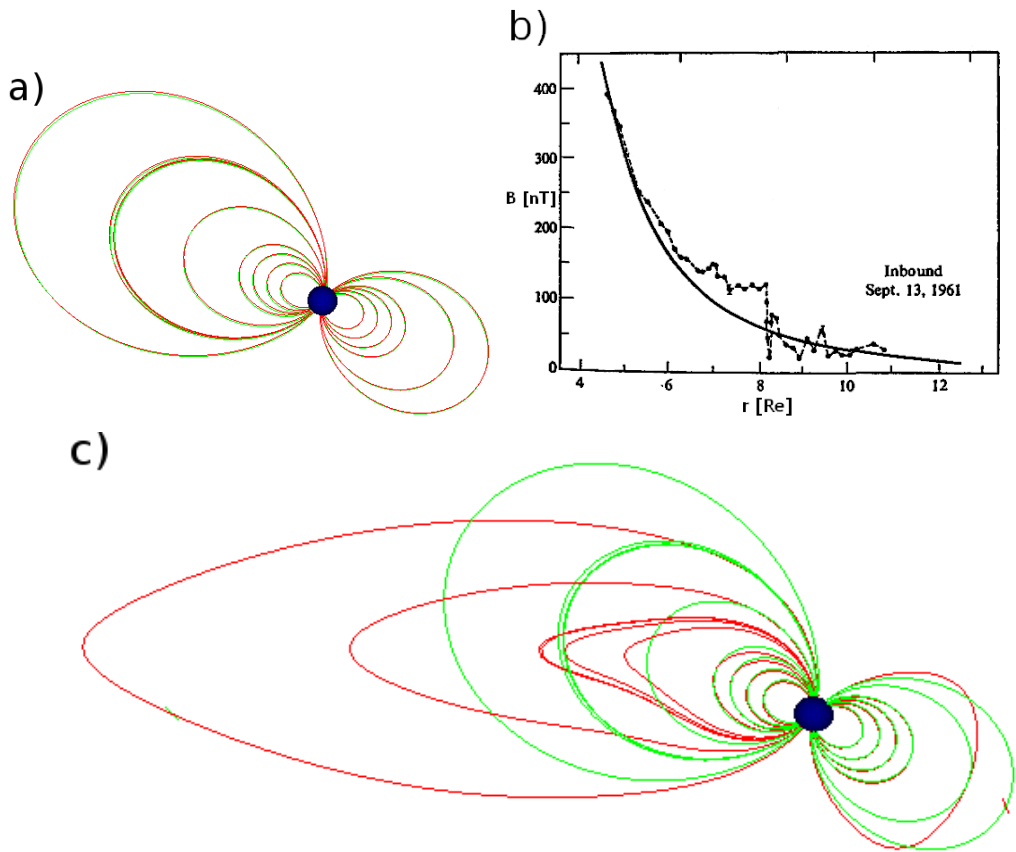

Figure 1. Geometry of dipole magnetic field centered at Earth center (green lines in a) and c)). a) IGRF model (red) b) magnetic field measurements by the spacecraft Explorer XII; the bold line is the spatial distribution of the dipole magnetic field strength. We see that until ~5$R_E$, the geomagnetic field is approximately dipolar (adapted from [5]). c) Field lines from the centered dipole model (green) together with those obtained from the model TSY01 (red). Earth size is at scale. A color version is available in the electronic version.

As expected, this model depends on the time of the day (due to the inclination of the geomagnetic axis respect to the Earth rotation axis) as well as on other effects linking the magnetosphere dynamics with the interplanetary conditions (e.g. compression of the magnetosphere due to Pdyn variations, or ring current excitations during geomagnetic storms).

The global variation of |**Bgeo**| at the Earth surface can be seen in figure 2 (upper panel). The secular evolution of the geomagnetic field is useful when comparing measurements involving long time periods. We consider this time evolution (using the TSY01 model) at Buenos Aires and found a rate change of -0.2% per year, as can be seen in figure 2 (bottom panel).

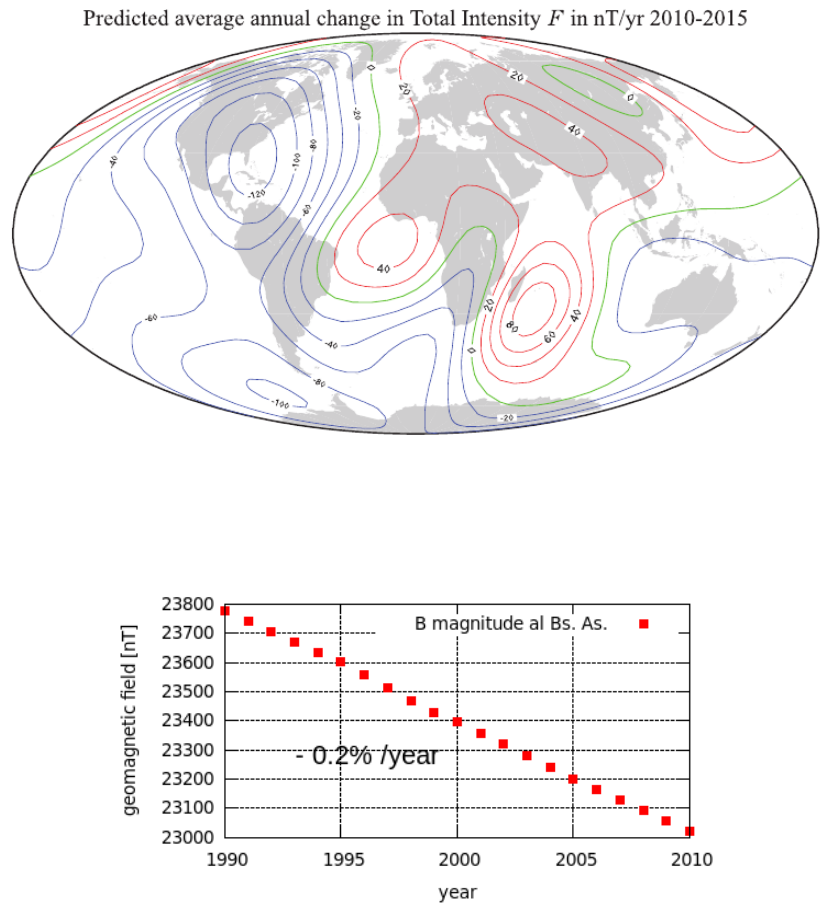

Figure 2. Upper panel: Map of the secular variation rate of |**Bgeo**| in the IGRF (2010 version) model (extracted from [3]). Bottom panel: Secular variation of |**Bgeo**|, evaluating the TSY01 model at Buenos Aires location. A color version is available in the electronic version.

## 2. Methodology

The transport of galactic cosmic rays (GCRs) implies several stages from the entry in the heliosphere until its detection at ground level. The last two stages are the entry and motion in to the magnetosphere and the interaction with the atmosphere, giving rise to the Extensive Air Shower. This work is focused in the transport from the entry to the magnetosphere until the top of the atmosphere; that is, the geomagnetic modulation on the primary particles.

In this work we define different configurations for the magnetic field in the magnetosphere in order to simulate the propagation of particles. When we use the model TSY01, we set the solar wind parameters as the typical ones for the interplanetary medium near Earth (e.g., solar wind dynamic pressure ~ 2nPa), and we only change the parameter Dst (see section 5) in order to change the level of the activity of the magnetosphere. Simulations are done for the geomagnetic configuration corresponding to January 1st, 2010, except for those studies of secular evolution.

## 2.1 Backtracking method

We are interested in the directions at which protons enter to the magnetosphere; that is, the asymptotic directions. Simulations are done for particles arriving at the top of the atmosphere above a given station (e.g. Buenos Aires or Marambio station), for which we denote its geocentric position as **L**.

Given that we have special interest in particles arriving at **L** position, we determine the backwards trajectory of a proton that arrived to **L**, with final momentum **p**. We do this by integrating the trajectory of an antiproton with initial position at **L** and initial momentum -**p** and we solve the equations of motion until either the trajectory length is greater than 100RE, or the particle reaches the magnetopause, or the trajectory is interrupted by the top of the atmosphere surface (that we consider at |**L**|=6390 km).

In the case that the trajectory length is greater than 100RE, two sub-cases are possible: either the particle was confined in the geomagnetic field or reached an asymptotic direction. This last case is due to the paraboloid shape of the magnetopause; so the boundary in the "tail" of the magnetosphere is not well defined.

In the case of an interrupted trajectory, it is considered that a proton with the given rigidity (**R=**c**p**/q, with c the speed of light and q the electric charge of the particle) cannot arrive to the position **L**. We called it an **allowed** trajectory if it is possible for the particle to arrive to **L**, and **forbidden** if otherwise.

So, we can determine a lower cutoff rigidity RL, above which there exist allowed trajectories with rigidities R> RL, for a given incidence direction. The value of RU is defined as the rigidity above which all trajectories are allowed (rigidities are such that R < RU). In practice there appears a mixed region (with rigidities R such that RL<R<RU), that correspond to allowed and forbidden trajectories, known as **penumbra** [6].

## 2.2 Transmittance function

In this work, we run the simulations using the MAGCOS code (http://cosray.unibe.ch/~laurent/magnetocosmics), which has a Geant4 platform.

In order to show the structure of the penumbra, we define the transmittance function setting it as 0 when the particle rigidity corresponds to an allowed trajectory, and as 1 if it is forbidden. In the following, we determine the structure of the transmittance function using several models of *Bgeo*, and the subsequent determination of an effective cutoff rigidity *Rc* (see Section 2.3).

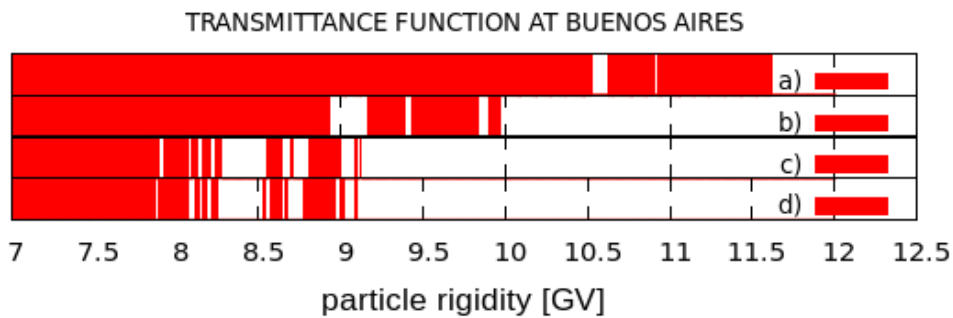

Figure 3. Transmittance function obtained with trajectory simulations using four Bgeo models: a) Centered Dipole b) Shifted Dipole c) IGRF (2010) d) IGRF+TSY01. Red corresponds to 0 (particle can reach the top of the atmosphere), and white to 1 (particle cannot reach it).

For vertical incidence (zenith=0°), the figure 3 shows the transmittance function for protons that arrive to Buenos Aires city (34.5°S, 58.4°). We used four models for *Bgeo*: Centered Dipole, Shifted Dipole (dipole center spatially shifted from Earth's center), IGRF2010 and TSY01; all of them with the real tilt angle (geomagnetic axis respect to the rotation one) [3]. The transmittances were determined with a step of ΔR=0.001GV in rigidity.

The most significant change occurs between the Centered Dipole model and the Shifted one. This is because the shift is in a direction almost opposite to Buenos Aires location on Earth's surface, causing a significant loss in the *Bgeo* strength.

## 2.3 Cutoff Rigidity at Bs. As. and Marambio

To obtain an effective cutoff rigidity Rc, we employ the definition (see [6]) Rc=$R_L$+N.ΔR (ΔR=0.001GV), where $R_L$ is the first low rigidity that does not bent back to Earth ("allowed" trajectory), and N is the number of allowed rigidities in the penumbra region. For vertical incidence, we obtain an effective cutoff rigidity at Buenos Aires of $R^{BA}_C$=8.41GV, and $R^{Mr}_C$=2.32GV at Marambio.

## 3 Validation test

In order to validate and compare our results, we reproduce some published similar simulations; we show two examples: a global distribution of the vertical rigidity cutoff and the transmittance function of the Newark NM location.

Simulations for particles arriving under vertical incidence at different location were done, obtaining the transmittance functions and the associated effective rigidity cutoff (Rc).

The values of Rc are color-encoded in the right panel of figure 4 (in a 5°x5° grid). We see that there is a 'cosmic ray equator' that roughly agrees with the geomagnetic equator [3]. A comparison with a similar map taken from the literature [7] (shown in left panel of figure 4) give us a positive test to our simulations.

On the other hand, we determine the transmittance function for the Newark NM location for arrivals under vertical incidence. In figure 7 we compare the results of these simulations with those of Smart D.F. et al (2000) [6], and we note that the $R_L$ and $R_U$ values are in a very good agreement; the RC values differ by less than 1%. This tiny difference might be due to the evaluation of the *Bgeo* model (IGRF) for a different day of the year, at a different time (in the publication, these details are not specified) or maybe due to the different integration method.

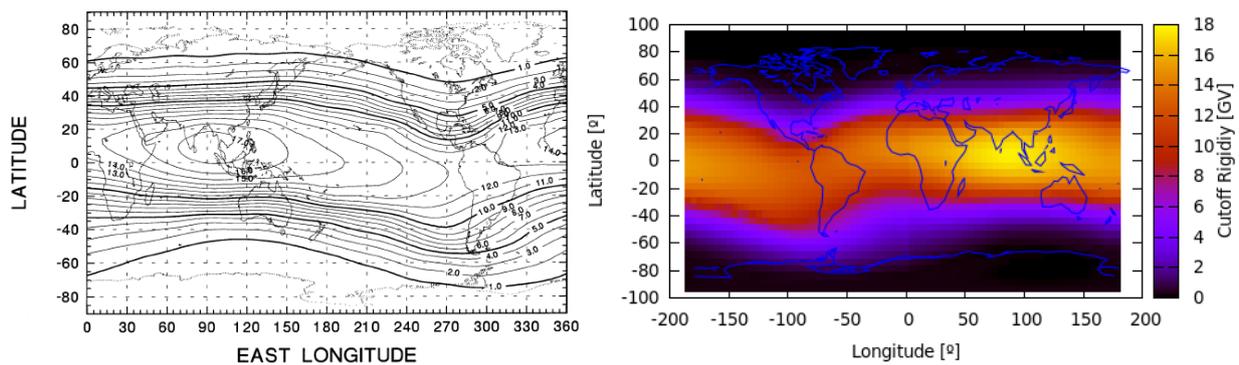

Figure 4. The vertical effective cutoff rigidity Rc in function of the position, in a 5°x5°grid, obtained by Smart D.F. et al (2008) [7] (left); and with our simulations with the IGRF model (right). A color version is available in the electronic version.

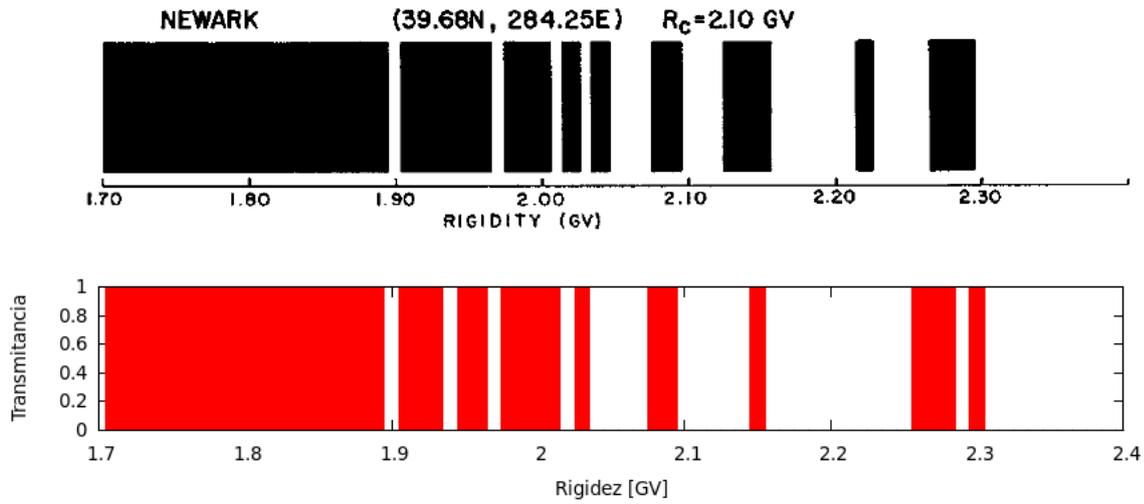

Figure 5. Transmittance function for Newark neutron monitor location obtained by Smart D.F. etal (2000) (up) and [6] and obtained with our simulations (down) as a validation test. The cutoff rigidity Rc of both results differ by less than 1%.

## 4 Determination of asymptotic directions

We determine asymptotic directions for particle rigidities above Rc for the location of each of the two stations; the results are shown in figure 6. We can see that as the particle rigidity decreases, the asymptotic trajectories get closer to the equator region.

From a detailed analysis of the simulations done it is possible to conclude that these particles are mostly deflected at heights lower than ~ 2 $R_E$, where the configuration of *Bgeo* is strongly dominated by a dipolar component. However, in next section we will see that during periods of geomagnetic storm, the non-dipolar component of *Bgeo* (produced mainly by magnetospheric electric currents) can significantly affect the trajectory of these particles.

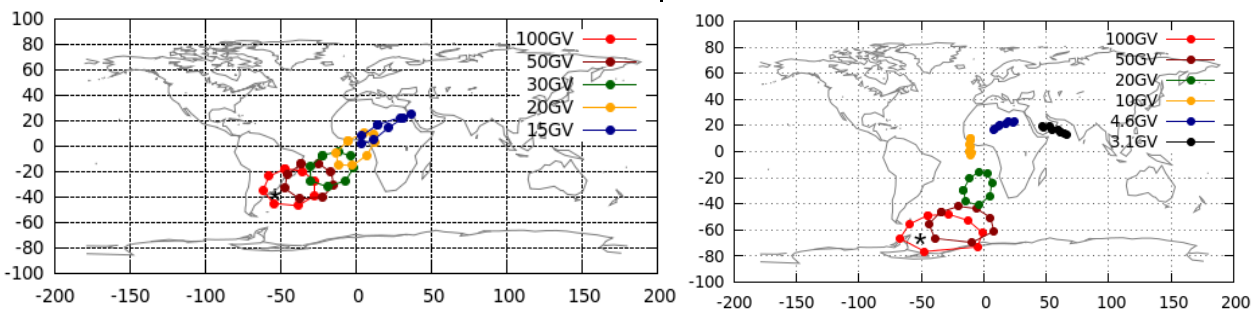

Figure 6. Asymptotic directions (projected on the Earth's surface) for 15° zenith incidence and eight incidence azimuth values (45°,90°,...,360°), for the two LAGO stations: Buenos Aires (left) and Marambio (right), using the IGRF2010+TSY01 model. The symbol * marks the position of particles arrival. A color version is available in the electronic version.

## 5 Effects of an active magnetosphere

The Dst index is a good proxy to determine the activity of the magnetosphere [12], and it is frequently used to quantify the intensity of the so-called geomagnetic storms, which are strong geomagnetic disturbances, which typically last ~10 hours [9].

We performed simulations to compute asymptotic directions for different geomagnetic conditions, considering quiet, intermediate, and active magnetospheric conditions.

In figure 7 we show the asymptotic directions of particles with different rigidities for different values of Dst (Dst=0, -100, -200, -300 and -400nT).

We note that the asymptotic directions tend to go westward as the storms get more and more intense; however, the main deflections keep occurring at altitudes from below 5RE, as happened during quiet conditions.

The quantitative information of interest in these results is the shift in longitude of theses asymptotic directions, with respect to the location of each station. This is of interest in order to determine the spatial CR flux anisotropy associated to the diurnal variation measured by neutron monitors [14].

From the transmittance functions we compute Rc for Buenos Aires and Marambio when Dst=0nT (calm period), and considering different periods of time. In the left panel of figure 7 it is possible to observe these results, which shows that $R^{BA}_c$ decreases roughly linearly at a rate of -0.04GV/year. While for Marambio, we find that $R^{Mr}_c$ decreases at a rate of -0.1GV/year; in agreement with the literature [7].

From evaluating the TSY01 model in the year 2010, we obtain the values of Rc as a function of Dst index (see right panel of figure 7).

The dependence of Rc with Dst is such that for active geomagnetic activity, lower energetic particles can reach ground level, compared to the quiet conditions. In particular, the decreasing trend is also linear with a rate of -0.001GV/nT at Buenos Aires, and -0.003GV/nT at Marambio.

During a geomagnetic storm, the day-night asymmetry is lightly emphasized. Figure 8 shows this effect for the two sites. It shows variations of Rc with local time along a day for different activity level of the magnetosphere (values of Dst from 0nT to -600nT). It is possible to see that as the storm gets more intense, there is an accentuated daily modulation and the average RC value decreases.

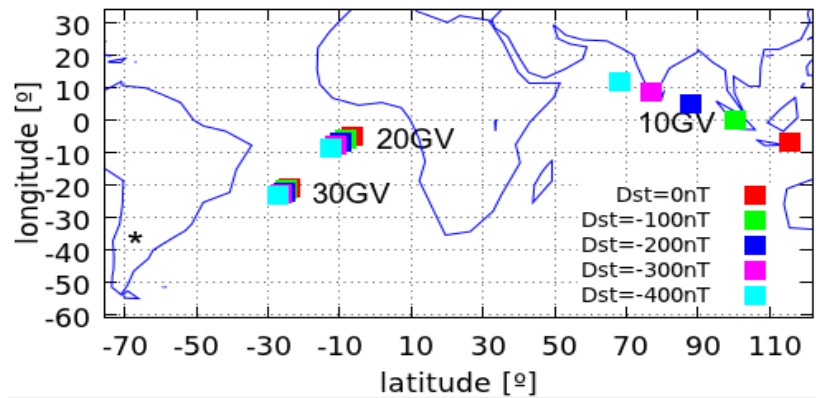

Figure 7. Asymptotic directions of proton trajectories (projected on Earth's surface) under geomagnetic storm conditions of Dst=0, -100, -200, -300 y -400nT for particle rigidities 10, 20 and 30 GV; all of them for vertical incidence (zenith=0) on Malargue. The asterisk symbol indicates Malargue location. A color version is available in the electronic version.

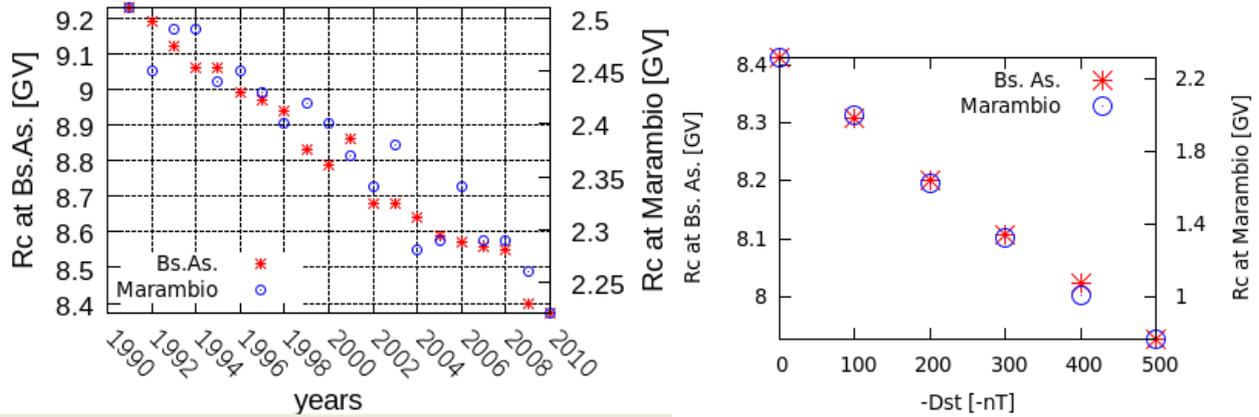

Figure 8. Left: Secular evolution of the cutoff rigidities $R^{BA}_C$ and $R^{Mr}_C$ along the last twenty years, with a lineal decreasing trend of -0.04GV/year for $R^{BA}_C$ and -0.1GV/year for $R^{Mr}_C$. Right: Evolution of the same cutoff rigidities as a function of the Dst index; the linear decreasing rate is of $\Delta Rc/\Delta Dst=-0.001$GV/nT at Buenos Aires and $\Delta Rc/\Delta Dst=-0.003$GV/nT at Marambio.

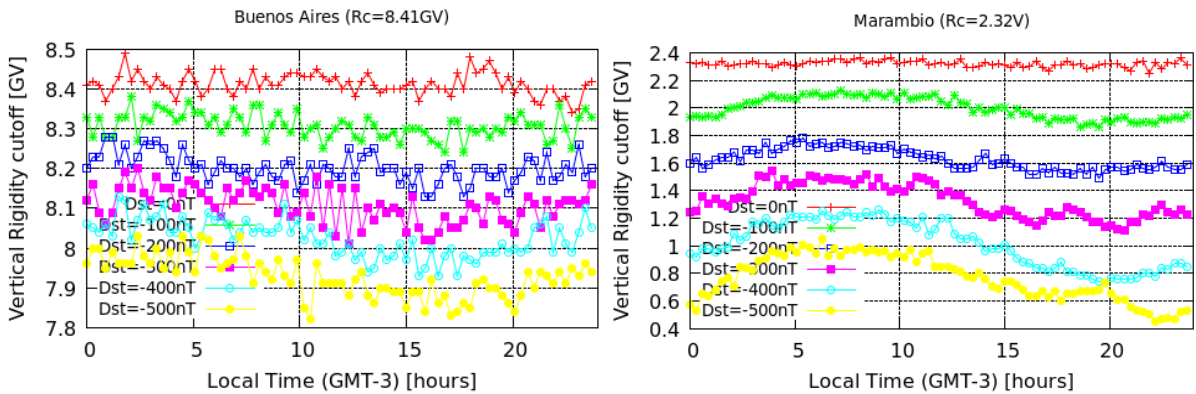

Figure 9. Effective rigidity for vertical incidence as a function of time, along one day for diferent storm conditions, at Buenos Aires (left) and Marambio (right) station. The modulation of RC is stronger as the storm gets more severe.

## 6 Summary and Conclusions

We have determined the cutoff rigidity for different incidence directions and different level of magnetospheric activity, using trajectories of a proton arriving to Buenos Aires and to Marambio (Antarctica), where two nodes of the net LAGO are planned to be built in the near future. We find $R^{BA}_C$=8.41±0.60 GV for Buenos Aires, and $R^{Mr}_C$=2.32±0.23GV for Marambio, where error values are given by the semi-width of the penumbra associated to each station.

We identified the asymptotic directions for particles arriving at these stations, and found that they do not change during the day. We determined the variation of Rc during one day for different geomagnetic storm conditions, and found a significant daily modulation for intense geomagnetic storms, in particular at Marambio.

We computed the variation of the transmittance function in the last 20 years, finding a -0.04GV/year decrease for $R^{BA}_c$ and -0.1GV/year for $R^{Mr}_c$. Simulations considering different values of the Dst index, show a decrease rate $\Delta Rc/\Delta Dst$ of -0.001GV/nT at Buenos Aires, and -0.003GV/nT at Marambio.

All these results can be used to analyze and interpret future observations from these LAGO stations, in particular studies of Forbush decrease, which generally are observed in coincidence with an enhanced level of geomagnetic activity.

## Acknowledgments

This work was partially supported by the Argentinean grants UBACyT-20020120100220 and PIP-11220090100825/10 (CONICET).

J.J.M.M. is a fellow of CONICET. S.D. is member of the Carrera del Investigador Científico, CONICET. S.D. acknowledges support from the Abdus Salam International Centre for Theoretical Physics (ICTP), as provided in the frame of his regular associateship. J.J.M.M. would like to thank to United Nations, the National Aeronautics and Space Administration (NASA) and the Japan Aerospace Exploration Agency (JAXA) for financial support to participate in the United Nations/Ecuador Workshop on the International Space Weather Initiative.

The authors would like to thank SEPTIMESS project for the MAGNETOCOSMICS code.

## References

[1] X. Bertou et al. The Large Aperture GRB Observatory. Proceed. of the 31st ICRC, 2009.

[2] J. J. Masas-Meza, X. Bertou and S. Dasso Geomagnetic effects on cosmic ray propagation for different conditions. Proceed. of the International Astronomical Union, 7:234-237, 2011.

[3] C. C. IAGA Working Group V-MOD: Finlay, S. Maus, C. D. Beggan, and et al. International geomagnetic reference field: the eleventh generation. Geophysical Journal International, 183(3):1216-1230, 2010.

[4] N. A. Tsyganenko. A model of the near magnetosphere with a dawn-dusk asymmetry. Journal of Geophysical Research, 107(A8), 2002.

[5] G. K. Parks. Westview Press, 2004.

[6] D. F. Smart, M. A. Shea, and E. O. Fluckiger. Magnetospheric models and trajectory computations. Space Science Reviews, 93:305, 333, 2000.

[7] D. F. Smart and M. A. Shea. World grid of calculated cosmic ray vertical cutoff rigidities for epoch 2000.0. Proceedings of the 30th International Cosmic Ray Conference, 1:737-740, 2008.

[8] http://ccmc.gsfc.nasa.gov/modelweb/sun/cuto_.html

[9] S. Dasso, D. Gómez, and C. H. Mandrini. Ring current decay rates of magnetic storm: A statistical study from 1957 to 1998. Journal of Geophysical Research, 107(A5):1059, 2002.

[10] W. D. Gonzalez, V. M. Vasyliunas, and et al. What is a geomagnetic storm? Journal of Geophysical Research, 99(A4):5771-5792, 1994.

[11] P. Pierre Auger collaboration, Abreu and et al. The Pierre Auger observatory Scaler Mode for the study of solar activity modulation of galactic cosmic rays. JINST, 6:P01003, 2011.

[12] W. D. Gonzalez, B. T. Tsurutani, and et al. The extreme magnetic storm of 1-2 september 1859. Journal of Geophysical Research, 108(A7):1268, 2003.

[13] B. Vargas, J.F. Valdés-Galicia. Calculation of the magnetic rigidity cutoff and the asymptotic cone of acceptance for the site of the Pierre Auger Observatory in Malargüe, Argentina Proceed. of the 32nd International Cosmic Ray Conference (ICRC2011), 10:245, 2011

[14] M. A. Pomerantz, and et al. The Cosmic Ray Solar Diurnal Anisotropy. Space Science Reviews, 12:75-130, 1971.